\documentstyle[11pt,twoside,paspconf]{article}
\def\etal{{\it et al.\ }}

\def\eg{{\it e.g.,\ }}

\def\ie{{\it i.e.,\ }}
\def\msol{\ifmmode {\>M_\odot}\else {$M_\odot$}\fi}
\def\zsol{\ifmmode {\>Z_\odot}\else {$Z_\odot$}\fi}
\def\cmsq{\ifmmode {\>{\rm\ cm}^2}\else {cm$^2$}\fi}
\def\psqcm{\ifmmode {\>{\rm cm}^{-2}}\else {cm$^{-2}$}\fi}
\def\pcubcm{\ifmmode {\>{\rm cm}^{-3}}\else {cm$^{-3}$}\fi}
\def\psqpc{\ifmmode {\>{\rm pc}^{-2}}\else {pc$^{-2}$}\fi}
\def\pcsq{\ifmmode {\>{\rm\ pc}^2}\else {pc$^2$}\fi}
\def\intensity{\ifmmode{{\rm erg\ cm}^{-2}{\rm\ s}^{-1}
      {\rm\ Hz}^{-1}{\rm\ sr}^{-1}}
      \else {erg cm$^{-2}$ s$^{-1}$ Hz$^{-1}$ sr$^{-1}$}\fi}
\def\iintensity{\ifmmode{{\rm erg\ cm}^{-2}{\rm\ s}^{-1} {\rm\
sr}^{-1}}\else
      {erg cm$^{-2}$ s$^{-1}$ sr$^{-1}$}\fi}
\def\flux{\ifmmode{{\rm erg\ cm}^{-2}{\rm\ s}^{-1}}\else {erg
cm$^{-2}$ s$^{-1}$}\fi}
\def\fluxdensity{\ifmmode{{\rm erg\ cm^{-2}\ s^{-1}\ Hz^{-1}}}\else {erg
cm$^{-2}$ s$^{-1}$ Hz$^{-1}$}\fi}
\def\phoflux{\ifmmode{{\rm phot\ cm}^{-2}{\rm\ s}^{-1}}\else {phot
cm$^{-2}$ s$^{-1}$}\fi}
\def\pintensity{\ifmmode{{\rm phot\ cm}^{-2}{\rm\ s}^{-1}
      {\rm\ Hz}^{-1}{\rm\ sr}^{-1}}
      \else {phot cm$^{-2}$ s$^{-1}$ Hz$^{-1}$ sr$^{-1}$}\fi}
\def\epuv{\ifmmode{{\rm erg\ cm}^{-3}{\rm\ s}^{-1}}\else {erg
cm$^{-3}$ s$^{-1}$}\fi}
\def\lum{\ifmmode{{\rm erg\ s}^{-1}}\else {erg s$^{-1}$}\fi}
\def\pholum{\ifmmode{{\rm phot\ s}^{-1}}\else {phot s$^{-1}$}\fi}
\def\nt{\ifmmode{{\rm cm^{-3}\ K}}\else {cm$^{-3}$ K}\fi}
\def\be{\begin{equation}}
\def\ee{\end{equation}}
\def\bea{\begin{eqnarray}}
\def\eea{\end{eqnarray}}
\def\beas{\begin{eqnarray*}}
\def\eeas{\end{eqnarray*}}
\def\Em{\ifmmode{{\cal E}_m}\else {{\cal E}$_m$}\fi}
\def\Tkev{\ifmmode{T_{\rm kev}}\else {$T_{\rm keV}$}\fi}
\def\emunit{\ifmmode{{\rm cm}^{-6}{\rm\ pc}}\else {cm$^{-6}$ pc}\fi}
\def\fesc{\ifmmode{f_{\rm esc}}\else {$f_{\rm esc}$}\fi}
\def\hubunits{\ifmmode {\>{\rm km\ s^{-1}\ Mpc^{-1}}}\else {km
s$^{-1}$ Mpc$^{-1}$}\fi}
\def\gta{\;\lower 0.5ex\hbox{$\buildrel > \over \sim\ $}}
\def\lta{\;\lower 0.5ex\hbox{$\buildrel < \over \sim\ $}}          
\def\kms{\ifmmode {\>{\rm km\ s}^{-1}}\else {km s$^{-1}$}\fi}

\newcommand{\T}{ {\scriptscriptstyle {\rm T}} }
\newcommand{\qrms}{ Q_{\scriptscriptstyle {\rm RMS}} }

\newcommand{\rmw}{ r_{\scriptscriptstyle {\rm MW}} }
\newcommand{\sz}{ {{\rm sz}} }

\begin{document}

\title{Warm Gas and Ionizing Photons in the Local Group}
\author{Philip R. Maloney}
\affil{Center for Astrophysics and Space Astronomy, University of Colorado}
\author{J.~Bland-Hawthorn}
\affil{Anglo-Australian Observatory}
\begin{abstract}
We investigate how much warm ($T\sim 10^6$ K) gas may exist in a Local
Group corona, and, in particular, whether such a corona could be
detected indirectly through its ionizing radiation. We calculate the
non-LTE cooling of warm, low-density gas and the resulting ionizing
photon emissivities for a range of metallicities. In principle a
corona can produce a photon flux which significantly exceeds the
cosmic ionizing background. However, the additional constraints which
can be imposed on intragroup gas rule out models which produce
detectable ionizing photon fluxes at the Milky Way's distance from the
Local Group barycenter.
\end{abstract}

\vspace{-0.05in}
\section{Introduction}
How much warm ($T\sim 10^6$ K) gas could there be in the Local Group?
This question dates back 40 years, to the classic Kahn \& Woltjer
(1959) ``timing mass'' paper; Kahn \& Woltjer suggested that the bulk
of the timing mass (which exceeded the visible stellar mass in the
Local Group by a large factor) was in the form of a warm ($T\simeq
10^6$ K), low-density ($n\sim 10^{-4}$ \pcubcm) plasma. This
identification was driven largely by the difficulty of observing such
gas directly. Although the mass of the Local Group is now believed to
be dominated by (possibly non-baryonic) dark matter, the question of
whether there is a substantial amount of warm gas present in the Local
Group has recurred in several different contexts. In the following
section we review the most recent suggestions for the presence of such
a Local Group ``corona''; in \S 3 we investigate whether such a corona
could be detected indirectly through its ionizing radiation. A fuller
discussion of this material is presented in Maloney \& Bland-Hawthorn
(1998a).

\section{A Local Group Corona?}
Suto \etal (1996) suggested that the measurement of the {\it COBE}
cosmic microwave background (CMB) quadrupole moment could be
significantly affected by the presence of a warm intragroup medium
centered on the barycenter of the Local Group: for an electron density
distribution of the form
\be
n_e(r) = n_o {r_o^2 \over r^2 + r_o^2},
\ee
where $n_o$ is the central density and $r_o$ is the core radius,
centered a distance $R$ from the Galaxy, the expected monopole and
quadrupole anisotropies are
\begin{eqnarray}
T_{0,\sz} &=& \pi \theta_o 
     \,\sigma_\T {kT \over mc^2} {n_o r_o^2 \over R} ,\\
T_{2,\sz} &=& {\sqrt{5}\pi \over 4}
\left[\theta_o - 3{r_o \over R} + 3\left({r_o \over R}\right)^2
\theta_o \right] 
     \,\sigma_\T {kT \over mc^2} {n_o r_o^2 \over R} ,
\end{eqnarray}
where $\theta_o \equiv \tan^{-1}(R/r_o)$. These equations can be
inverted to get
\begin{equation}
n_o r_o T_{\rm keV} < 7.4\times10^{21}\theta_o^{-1}{R\over r_o}
 \left({y \over 1.5\times10^{-5}}\right)\;\psqcm. 
\end{equation}
\be
n_o r_o T_{\rm keV} < 1.6\times 10^{20} Q_{{\mu\rm K}}{R\over r_o}
\left[\theta_o-3\left({r_o\over R}\right)+3\theta_o\left({r_o\over
R}\right)^2\right]^{-1}\;\psqcm
\ee
where the upper limit to the Compton $y-$parameter is $|y|=
T_{0,\sz}/2 < 1.5\times10^{-5}$ (95\% confidence level: Fixsen \etal
1996), while the rms quadrupole amplitude $\qrms=10^{-6}Q_{{\mu\rm
K}}$ K; the observed value $Q_{{\mu\rm K}}\approx 6$ (\eg Bennett
\etal 1996). Suto \etal argued that a LG corona which satisfied 
the inequality (4) could still significantly impact the CMB quadrupole
measurement. 

This conclusion was immediately contradicted by Banday \& Gorski
(1996), who demonstrated that the {\it COBE} data show no evidence for
such a distortion, and by Pildis \& McGaugh (1996), who pointed out
that X-ray observations of poor groups, when fitted with density
profiles of the form (1), typically obey $n_o r_o\Tkev \lta 3\times
10^{20}$ \psqcm, with $\Tkev\sim 1$. Furthermore, most {\it detected}
groups are dominated by ellipticals; the upper limits to $n_o
r_o\Tkev$ for spiral-rich groups are usually an order of magnitude
smaller. Thus, although the {\it COBE} constraints on a LG corona are
quite weak, analogy with similar (indeed, usually richer)
poor groups suggests that the LG is unlikely to possess a significant
gaseous halo.

However, Mulchaey \etal (1996), in an X-ray study of poor groups,
suggested that spiral-rich groups might be undetected not because of
an absence of gas, but because the gas is too cold to detect:
spiral-dominated groups tend to have lower velocity dispersions,
implying virial temperatures of only $T_{\rm vir}\sim 0.2$ keV. The
soft X-ray emission from gas at such temperatures is extremely
difficult to detect. If the product $n_o r_o\Tkev$ is similar
to that in detected groups, the gas mass could be substantial: the
mass within radius $r$ is
\be
M(r)\sim 7\times10^{11}\left(r_o\over 100\;{\rm
kpc}\right)^2\left({r\over r_o}\right)\left({n_o r_o\Tkev\over 10^{20}
\;\psqcm}\right)\left({0.2\over\Tkev}\right)\;\msol
\ee
for $r/r_o\gta$a few. (Note, however, that X-ray-detected poor groups
usually have $M_{\rm gas}/M_{\rm grav}\sim 0.1$, which limits the
total gas mass.) In fact, Wang \&
McCray (1993) found evidence from {\it ROSAT} observations for a
diffuse thermal component, with $\Tkev\sim 0.2$ and $n_e\sim
10^{-2}x^{-1/2}\pcubcm$, where $x$ is the line-of-sight extent of the
emitting region in kpc; the inferred electron density is smaller by a
factor $\sim3$ for solar metallicity rather than primordial gas. Since
only the emission measure \Em\ is obtainable from the data, the
spatial extent is unknown.

Blitz \etal (1998) have suggested that most HVCs are remnants of the
formation of the Local Group, associated with continuing infall onto
the LG. (See also Blitz, this volume.) If true, some fraction of these
clouds will collide in the vicinity of the LG barycenter, and shock up
to the virial temperature of the LG, $T\sim 2\times 10^6$ K, leading
to the formation of a warm intragroup medium.

In the next section we consider whether the ionizing photon flux from
such a corona could be detectable, and the additional constraints
which can at present be imposed on a LG halo.

\section{Ionizing Photons from a Local Group Corona}
We assume a density distribution of the form (1), with an outer
boundary $r_b$ and a temperature
$\Tkev\simeq 0.2$, the virial temperature of the LG. To calculate the 
emission and the ionizing photon flux from the gas, we have used the
photoionization/shock code MAPPINGS, kindly provided by Ralph
Sutherland. We consider metallicities $Z=0.01$, $0.1$, and
$0.3\zsol$. As long as $r_b\gta$a few $r_o$, the ionizing photon flux
normally incident on a plane-parallel gas layer is
\be
\phi_i(r) \approx 10^4 n^2_{-3} r_{100} \left({\xi_i\over 10^{-14}}\right)
{\left[0.8+1.3(r/r_o)^{1.35}\right]\over(1+r^2/r_o^2)^{1.5}}\ {\rm
cm^{-2}\; s^{-1}}
\end{equation}
where the central density $n_o=10^{-3}n_{-3}\pcubcm$ and the core
radius $r_o=100 r_{100}$ kpc. For the temperatures and metallicities
of interest, the integrated photon emissivity $\xi_i$ generally lies
between $3\times 10^{-15} - 3\times 10^{-14}\;\pcubcm$ s$^{-1}$
sr$^{-1}$. Poor groups show a very broad range of core radii, from
tens to hundreds of kpc (Mulchaey \etal 1996), and typical central
densities $n_o\sim\;$a few$\,\times 10^{-3}\pcubcm$ (Pildis \& McGaugh
1996). For these parameters, the photon fluxes predicted by equation
(7) are quite substantial. (For reference, the cosmic ionizing
background [the integrated contribution from AGN and galaxies] is
probably $\phi_{\rm i,cos}\sim 10^4$ \phoflux: Maloney \&
Bland-Hawthorn 1998b.) In Figure 1, we show (shaded in gray) the range
of corona parameters ($r_o, n_o$) for which the resulting ionizing
photon flux is between $\phi_i=10^4$ and $\phi_i=10^5$ \phoflux, for
radial offsets $r=0$ (upper region) and $r=350$ kpc (lower region).
The latter is the Galaxy's assumed radial distance $\rmw$ from the
center of the LG. However, there are additional constraints which we
can place on a LG corona, which rule out a significant (\ie $\phi_i >
\phi_{\rm i,cos}$) contribution to the ionizing photon flux at $\rmw$:
\begin{itemize}
\item The assumption that any LG intragroup medium is typical of poor
groups constrains the product $n_o r_o \lta
1.5\times 10^{21}\psqcm$, for an assumed temperature of $\Tkev\approx
0.2$. This is plotted as the short-dashed line in Figure 1. Any corona
which is not unusually rich must lie to the left of this line. This
requirement alone rules out any significant contribution to $\phi_i$
at $\rmw$. 
\item  The timing mass -- obviously, the mass of the corona cannot
exceed the mass of the LG. There are still come uncertainties in the
determination of the timing mass (see Zaritsky 1994, and this volume,
for a summary). We have adopted a value for the timing mass of
$M_T=5\times 10^{12}$ \msol\ within $r=1$ Mpc of the LG center. The
mass in the corona is given by 
\be
M_c(r)=4.3\times10^{11} n_{-3} r_{100}^3 x\left[1-{1\over
x}\tan^{-1}(x)\right] \msol
\ee
where $x=r/r_o$, and for Figure 1 we have calculated the corona mass
out to $r=1$ Mpc. The timing mass constraint is shown as the solid
line in Figure 1. As plotted, this is barely more restrictive than the
{\it COBE} quadrupole limit (shown as the long-dashed line) and is
only more restrictive than the assumption of a typical intragroup
medium for large core radius. However, the true timing mass limit is
considerably more stringent than this, as we must subtract off the
masses of M31 and the Milky Way.
\item Limits on the actual electron density at $r\sim \rmw$. These
constraints come from two sources. First, observations of dispersion
measures toward pulsars in the LMC and a distant globular cluster, NGC
5024 (Taylor, Manchester \& Lyne 1993) require a mean $n_{-3}\sim
1$. However, most of this column must be contributed by the Reynolds
layer, so only a small fraction (probably $\lta 10\%$) can be due to a
LG corona. Secondly, an average density of no more than $n_{-3}\sim
0.1$ is allowed by models of the Magellanic Stream; otherwise, the
Stream clouds would be plunging nearly radially into the Galaxy (Moore
\& Davis 1994). The hatched region in Figure 1 indicates the portion
of $(r_o,n_o)$ space in which $n_e(\rmw)\le 10^{-4}$ \pcubcm.
\end{itemize}

\begin{figure}[!htb]
\plotfiddle{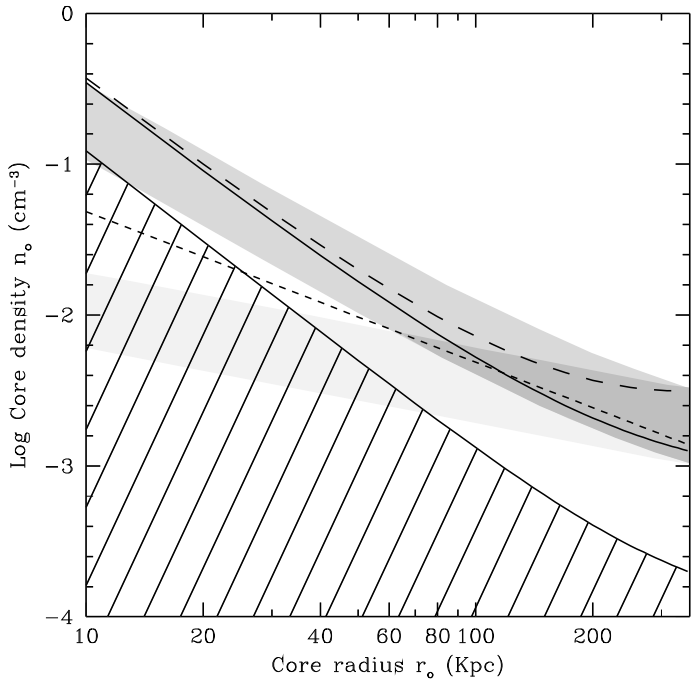}{80truemm}{0}{65}{65}{-200}{-160}
\caption{Constraints on a Local Group corona in the $(r_o,n_o)$ plane.
Coronae within the gray-shaded regions produce ionizing photon fluxes
between $\phi_i=10^5$ and $10^4$ \phoflux \ (upper and lower edges)
at radii $r=0$ (lower region) and $r=350$ kpc (upper
region) with respect to the LG center. The long-dashed line is the
{\it COBE} quadrupole constraint, the short-dashed line assumes the LG
medium is ``typical'', the solid line is the timing mass constraint,
and the hatched region satisfies $n_e\le 10^{-4}\pcubcm$ at $r=350$
kpc. See text for discussion.}
\end{figure}
The most stringent constraints come from the limits on the local value
of $n_e$ and from the timing mass (when realistic contributions from
the Galaxy and M31 are included). Although it is possible to
circumvent these restrictions to some extent by assuming the gas is
highly clumpy, it is not possible to evade the soft X-ray
determinations from Wang \& McCray (1993) in this fashion, since this
is a measurement of the emission measure. If the $\Tkev\approx 0.2$
component inferred by Wang \& McCray is extended on the scale of the
local group, then the mean density is at most $n_{-3}\sim 0.3$, not
far above the hatched region in Figure 1. These constraints rule out
any significant ionizing photon flux at $r\sim \rmw$ from a Local
Group corona: if the core density $n_o$ is high, then $r_o$ must be
small, while if $r_o$ is large, $n_o$ must be low. LG coronae in the
allowed region of $(r_o,n_o)-$space could produce ionizing fluxes much
larger than the cosmic background, but only on scales of a few tens of
kpc at best. An allowed LG medium could still contain a substantial
amount of mass; the direct limits on such a medium have yet to improve
on the values suggested by Kahn \& Woltjer (1959).

\acknowledgments
PRM acknowledges support from the Astrophysical Theory Program
through NASA grant NAG5-4061. We are grateful to Simon White for
comments and platypus spotting.


\end{document}